\documentclass[prl,twocolumn,aps,amsmath,amssymb,showpacs,superscriptaddress]{revtex4-1}

\usepackage{graphicx}% Include figure files
\usepackage{bm}
\usepackage[english]{babel}
\usepackage{amsmath} %some extra symbols
\usepackage{amsfonts} %some extra symbols
\usepackage{amssymb} %some extra symbols
\usepackage{color}

\begin{document}

\title{Cooper Pair Splitting by means of Graphene Quantum Dots}

\author{Z. B. Tan}
\affiliation{O.V. Lounasmaa Laboratory, Aalto University, P.O. Box 15100, FI-00076 AALTO, Finland}
\author{D. Cox}
\affiliation{O.V. Lounasmaa Laboratory, Aalto University, P.O. Box 15100, FI-00076 AALTO, Finland}
\author{T. Nieminen}
\affiliation{O.V. Lounasmaa Laboratory, Aalto University, P.O. Box 15100, FI-00076 AALTO, Finland}
\author{P. L\"ahteenm\"aki}
\affiliation{O.V. Lounasmaa Laboratory, Aalto University, P.O. Box 15100, FI-00076 AALTO, Finland}
\author{D. Golubev}
\affiliation{O.V. Lounasmaa Laboratory, Aalto University, P.O. Box 15100, FI-00076 AALTO, Finland}
\author{G.\,B.\,Lesovik}
\affiliation{L.D.~Landau Institute for Theoretical Physics RAS, Chernogolovka, 142432, Moscow Region, Russia}
\author{P. J. Hakonen}\email[Corresponding author: pertti.hakonen@aalto.fi]{}
\affiliation{O.V. Lounasmaa Laboratory, Aalto University, P.O. Box 15100, FI-00076 AALTO, Finland}

\begin{abstract}

Split Cooper pair is a natural source for entangled electrons which is a basic ingredient for quantum information in solid state. We report an experiment on a superconductor-graphene double quantum dot (QD) system, in which we observe Cooper pair splitting (CPS) up to a CPS  efficiency of $\sim 10\%$. With bias on both QDs,  we are able to detect a positive conductance correlation across the two distinctly decoupled QDs. Furthermore, with bias only on one QD, CPS and elastic co-tunneling can be distinguished by tuning the  energy levels of the QDs to be asymmetric or symmetric with respect to the Fermi level in the superconductor.

\end{abstract}


\maketitle

%\section{Introduction}

A Cooper pair, splitting from a superconductor into two different normal metal terminals \cite{lesovik2001,Recher2001}, is a natural source of non-local entangled electrons, which is an essential resource for quantum information processing \cite{loss1998}. During the past decade, many efforts have been made to split Cooper pairs into metal \cite{beckmann2004,russo2005,cadden2009,wei2010}, InAs nanowire \cite{hofstetter2010,hofstetter2011,Das2012} and carbon nanotube \cite{herrmann2010,schindele2012}. In a normal-superconductor-normal (NSN) type of metallic structure, evidence of entangled pairs has been reported using combined conductance and noise correlation measurements in two SN junctions \cite{wei2010}. By replacing the normal metal with QDs, the splitter concept was essentially upgraded and efficient CPS was demonstrated by manipulating the energy levels of the two QDs \cite{hofstetter2010}. Subsequently,  the splitting efficiency was improved up to 90\%  on a carbon nanotube QD device \cite{schindele2012}. However, since both QDs were on the same nanowire or -tube, applying bias independently on both sides remained impossible.

Graphene, a single sheet of carbon atoms, has unique physical properties as a Dirac Fermion system. Besides nearly ballistic conductance with huge mean free paths, spin-orbit coupling in graphene is expected to be weak which implies long spin coherence times. According to theory,  production of entangled electrons  by CPS works well in graphene \cite{cayssol2008,benjamin2008}. Due to the unique properties of graphene, it is also possible that a Majorana fermion, a fermion which is its own antiparticle, would be stabilized in the superconductor-graphene junction \cite{schaffer2012,chamon2012}. So far, distinct bipolar supercurrents have been observed in superconductor-graphene-superconductor junctions \cite{heersche2007}. In addition,  sharp Andreev bound states have been distinguished in a graphene quantum dot by making a good tunnel barrier between the superconductor and the graphene sheet \cite{dirks2011}. The splitting of a Cooper pair into graphene QDs might involve Majorana bound states,  which would yield an extra dimension for the CPS in this system.

CPS in graphene, however, has not been achieved so far. A major concern in this issue has been the irregular graphene edge, which has been proven to play an important role in the electrical transport of graphene ribbons \cite{guttinger2012}. Whether edges will strongly degrade the CPS in patterned graphene is an open question. Here, we report the first demonstration of CPS in graphene using patterned quantum dots. And this time, contrary to nanowire experiments, we are able to tune independently the bias of the two QDs. Using conductance correlations and current splitting between separately biased output terminals while tuning over a resonance level, we obtain a splitting efficiency of 10\%, which is clearly larger than expected by the current theories.
\begin{figure}[htbp]
\includegraphics[width=0.9\linewidth]{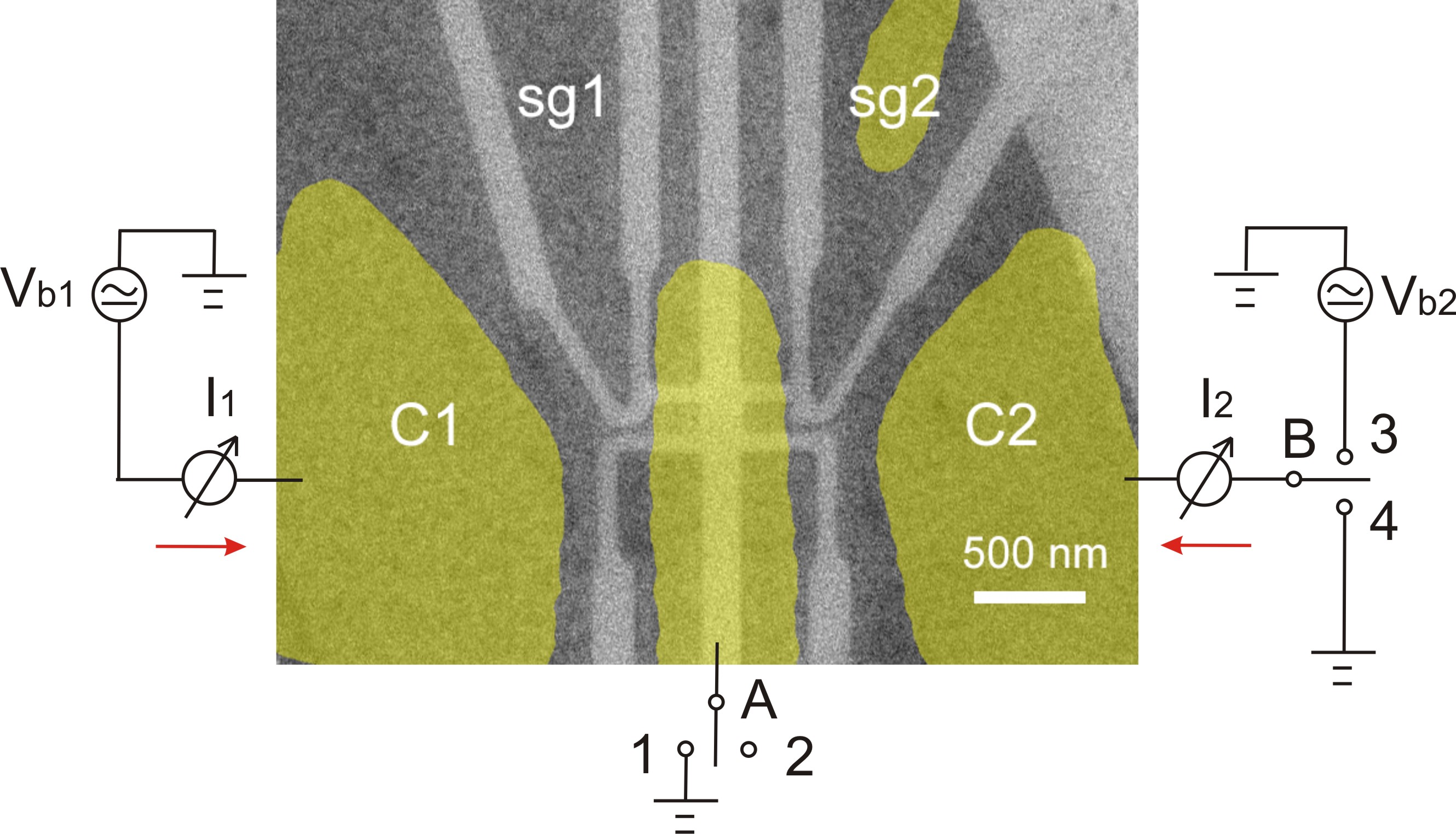}
\caption{ False color scanning electron micrograph of our graphene sample, where the graphene and Ti/Al contacts are indicated in dark gray and yellow, respectively. The light gray is uncovered $\mathrm{SiO_2}$ substrate. Two rectangle-shaped  QDs and two side gates were defined by EBL and oxygen plasma etching. Bias is separately applied on the two QDs. Switches A  and B are used to indicate the different biasing configurations.}
\end{figure}

We used mechanical exfoliation to extract graphene onto a highly doped silicon substrate coated with 267 nm of thermal silicon dioxide. The graphene device consists of  two spatially separated QDs with size about $200\;\mathrm{nm} \times 150\;\mathrm{nm}$, two graphene ribbons about $160\;\mathrm{nm} \times 50\;\mathrm{nm}$ connecting the QDs to two large output terminals,  and two side gates for tuning the dots, which were all defined using electron-beam lithography (EBL), followed by oxygen plasma etching via a PMMA mask (see Fig. 1). A second EBL step was used to define the metallic leads (in yellow) for connecting all the measurement electrodes. After development, 5/50 nm Ti/Al was evaporated onto the sample using electron-beam evaporation, followed by a lift-off process. The advantage of our device geometry for CPS is that the two QDs are on different pieces of graphene and separated by a gap of $r=180$ nm. The only electrical connection between the dots is through the common Al lead in the middle. This allows us to separately tune the bias on both output electrodes. In a diffusive superconductor, the coherence length $\xi = \sqrt{ \xi _{0}l } $ is estimated to be about 180 nm \cite{Feinberg2003}, taking the coherence length of a clean Al $\xi _{0}$ as 1600 nm and the mean free path $l=20$ nm deduced from the measured resistance. In principle, one might reach higher splitting efficiency with smaller $r$. The reason for having such a large gap $r=180$ nm between the QDs is to suppress the capacitive coupling between the two QDs  while keeping $r$ still around $\xi$. The employed separation appears to be an appropriate compromise, as other measured devices with $r \sim 100$ nm did not improve the splitting efficiency significantly.

Our measurements were made using a dry dilution refrigerator with a base temperature around 50 mK. Standard lock-in techniques were employed to measure differential conductance, normally using a 20 $\mu$V ac bias voltage at 7 Hz frequency. Dc voltage bias was only used for measuring the superconducting gap in Fig. 2(b) and 2(c).  Currents $I_1$ and $I_2$ were measured on both QDs with the red arrow denoting the positive direction (see Fig. 1). Several configurations were measured, which were obtained by operating the switches A and B in Fig. 1. Fig. 2(a) displays the conductance of both dots, $g_1$ and $g_2$, as a function of the back gate voltage with the Al lead grounded (with the switch A at position 1 and  B at position  3). The conductance variation is found to be quasiperiodic, with fluctuations caused by the level spacing statistics of the quantum dots as well as due to the ribbon constrictions between the dot and the large piece of graphene surrounding contacts C1 and C2. Close to $V_{g}=0$, the conductance is strongly suppressed, which indicates the Dirac point of the graphene sample is around zero gate voltage.
\begin{figure}[htbp]
\includegraphics[width=0.9\linewidth]{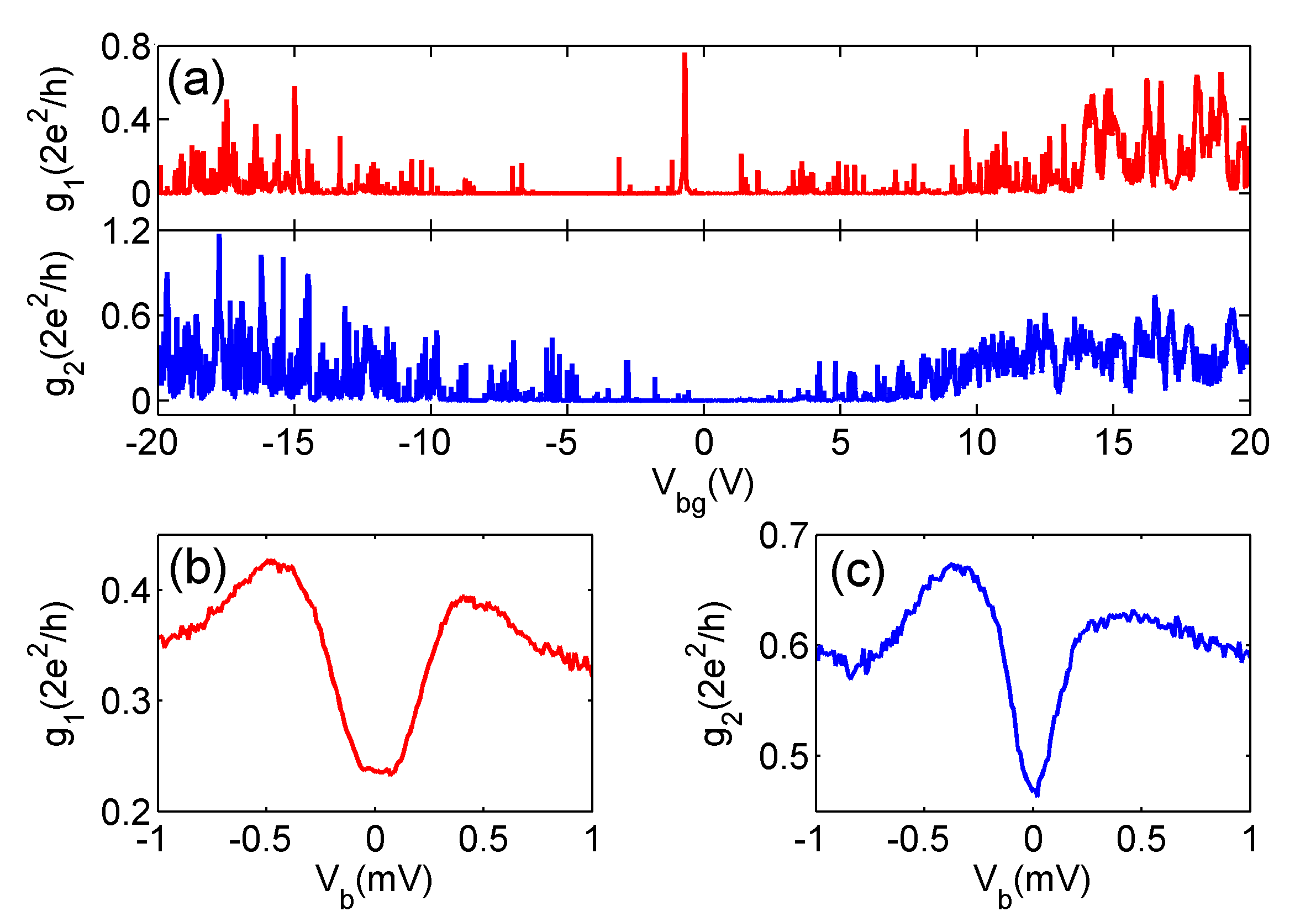}
{\caption{(a) Differential conductance of the two QDs $g_1=dI_1/dV_1$ and $g_2=dI_2/dV_2$ as a function of the back gate voltage with the superconducting terminal grounded. (b) Differential conductance $g_1$ as a function of bias voltage at $V_{bg}=20$ V. (c) Differential conductance $g_2$ as a function of bias voltage at $V_{bg}=20$ V.}}
\end{figure}

Figs. 2(b) and 2(c) reflect the influence of the superconducting gap of Al-graphene junction on the low bias conductance, measured on both sides. The differential conductance is measured as a function of bias voltage and, normally, the density of states of Al would lead to peaks at $V = \pm 0.18$ mV, or even lower if the Ti layer has reduced the value of the gap. However, due to  the geometry of our sample, $g_1$ and $g_2$ not only include Al-graphene junction resistance, but also the resistance of the graphene nanoribbon (GNR) constriction and even the large outer part of graphene and the contact resistance C1 and C2. Consequently, the bias only partly drops on the Al-graphene junction because the electron is apparently unable to tunnel through the GNR constriction in one shot with tunneling in the Al-graphene junction. The electron will be scattered by the irregular graphene edge or charge impurities in the silicon dioxide. Hence, the gap we measure from Figs. 2 (b) and (c) is clearly larger than the expected superconducting gap \cite{NOTE}. %It is noted that due to low contact resistance C1 and C2, the voltage droping on C1 and C2 is very small, so resistance C1 and C2 are roughly constant within measuring bias.
By analyzing the measured quantum dot transport, we obtain for the addition energy  $\Delta \mu \simeq 3$ meV and the width of the quantum dot resonance peaks $\Gamma = 0.1-0.2$ meV. Even though this $\Gamma$ is close in value to the gap energy, the quantum dots are functioning as energy filters in the splitting process (see below).

To determine the coupling between the two QDs, we measured the conductance $g$ of two dots in series as a function of side gate voltages $V_{sg1}$ and  $V_{sg2}$ (see Fig. 3) by keeping the Al electrode floating and forcing the measurement current to pass through both of the QDs  (the switch A at position 2, and B at position 4)  \cite{wiel2003}. If the coupling between the two QDs is very strong, the two dots can basically be considered as one dot and the side gates will tune the two QDs as a whole. If the coupling between the two QDs is intermediate, $g(V_{sg1},V_{sg2})$ will show hexagonal domains. If the coupling between the two QDs is very weak, the QD will be  tuned only by its individual side gate and the incoherent conductance of the two QDs in series $g(V_{sg1},V_{sg2})=g_{1}(V_{sg1})g_{2}(V_{sg2})/(g_{1}(V_{sg1})+g_{2}(V_{sg2}))$. The pattern in Fig. 3 shows that the two QDs are  tuned almost independently by their respective side gate, i.e.  very weak coupling between the QDs, which indicates small likelihood for nonlocal effects due to coupling.
\begin{figure}[htbp]
\includegraphics[width=0.9\linewidth]{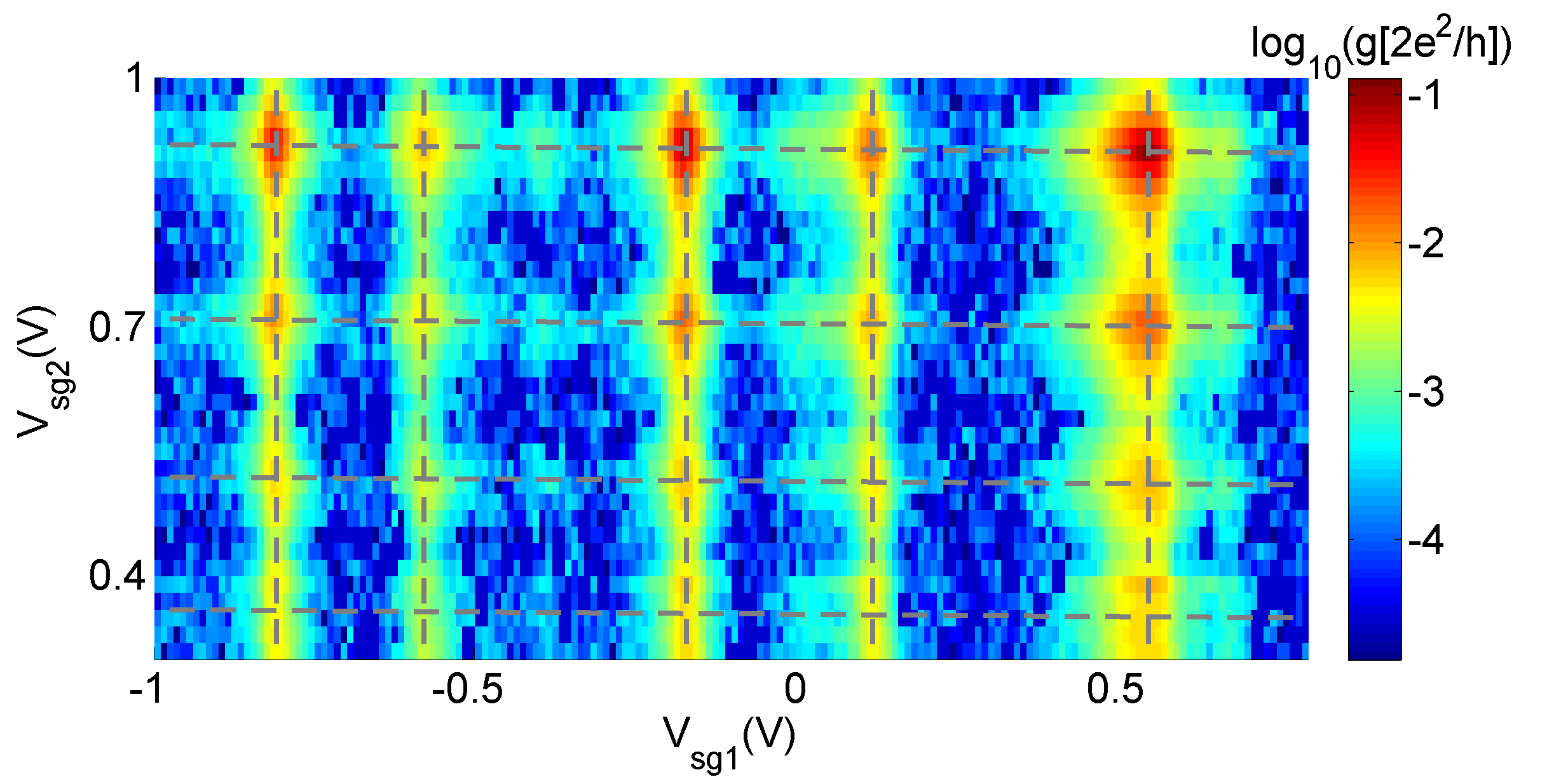}
{\caption{Differential conductance $g$ of the two QDs in series  as a function of $V_{sg1}$ and $V_{sg2}$ at $V_{bg}=5\;V$. The pattern shows that there is almost no coupling between the two QDs. The color bar on the right denotes the logarithm of the scaled conductance in units of $2e^2/h$.}}
\end{figure}

Evidence of CPS is shown in Fig. 4. The differential conductance of both QDs, $g_1$ and $g_2$, was measured simultaneously with equally large ac bias having the Al lead grounded (the switch A at position 1 and B at 3). By fixing $V_{bg} = 5$ V and $V_{sg2} = 0$ V, QD2 can be prepared in a conducting, resonant state. When the energy level of QD1, tuned by $V_{sg1}$, crosses the Fermi surface, electrons will tunnel easily through QD1, leading to conductance peaks in $g_{1}$. Because of the high addition energy of this QD, transport of both electrons of the Cooper pair to QD1 will be suppressed. So while one electron of the Cooper pair enters QD1, it is more likely that the other electron will go to QD2. This will lead to a conductance increase in $g_2$. Consequently, for each conductance peak in $g_{1}$, there will be a corresponding conductance peak in $g_{2}$. This nonlocal positive conductance correlation is regarded as evidence of CPS \cite{hofstetter2010,hofstetter2011,Das2012}.
\begin{figure}[htbp]
\includegraphics[width=0.9\linewidth]{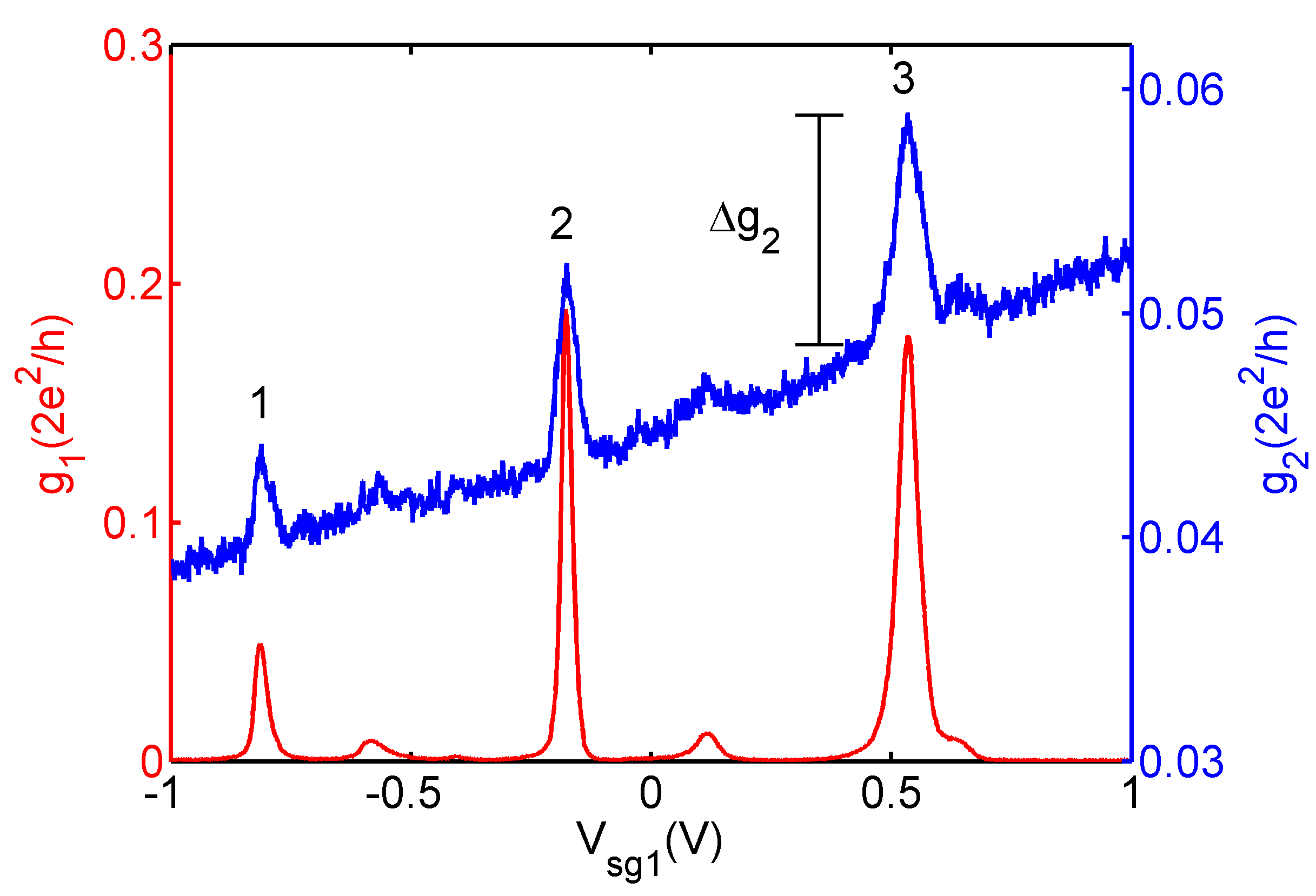}
{\caption{Conductance correlation between $g_1$ (red) and $g_2$ (blue) as a function of $V_{sg1}$ with $V_{bg} = 5$ V and $V_{sg2} = 0$. Here both QDs were ac biased at equal amplitude $V_{b1}=V_{b2}$, while the injector was grounded.}}
\end{figure}

Besides the capacitive-coupling-induced nonlocal effects, which were found to be small in our sample, the wire resistance in the common lead \cite{hofstetter2010} might produce negative nonlocal conductance which is distinct from the positive nonlocal conductance in CPS. However, the wire resistance is only about 15 Ohm in our setup, so its contribution is also very small. Thus, we can conclude that the nonlocal positive conductance correlation between $g_2$ and $g_1$ in Fig. 4 is due to CPS.

By defining CPS efficiency as the ratio of the number of split Cooper pairs to the total Cooper pairs involved in the transport,
%(the total number of Cooper pairs includes split Cooper pairs, Cooper pairs only transported to QD1 and Cooper pairs only transported to QD2)
we can write the CPS efficiency as \begin{eqnarray}
\begin{split}
\eta &= \frac{2\Delta g_{2}}{g_{1}+g_{2}}
\end{split}
\end{eqnarray}
where
%$g_{1}$ is the differential conductance in QD1, $g_{2}$ is the differential conductance in QD2 and
$\Delta g_2$ refers to the conductance increase in QD2 due to CPS. The CPS efficiency at peaks 1, 2 and 3 in Fig. 4 amounts to $10\%$, $7.5\%$, and $8.9\%$, respectively. Here, the quasiparticle tunneling rates in $g_{1}$ and $g_{2}$ are included in the base values, which lowers the achieved splitting efficiency.

According to the theory, the non-local conductance in our setup, $g_{12}=\partial I_1/\partial V_2\approx \Delta g_2$,
is given by the difference between CPS and elastic cotunneling (EC) contributions,
$g_{12}=G_{CPS}-G_{EC}$. In case of two point-like junctions attached to a disordered 2d superconducting
film with the sheet resistance $R_{\Box}$, these contributions read \cite{Feinberg2003,Golubev2009}
\begin{equation}
\begin{split}
G_{CPS} = \frac{R_{\Box}}{8}K_0\left(\frac{\sqrt{2}r}{\xi}\right)
\int \frac{G_1(E)G_2(-E)}{4T\cosh^2(E/2T)} dE \\
G_{EC} = \frac{R_{\Box}}{8}K_0\left(\frac{\sqrt{2}r}{\xi}\right)
\int \frac{G_1(E)G_2(E)}{4T\cosh^2(E/2T)} dE
\end{split} \label{cond}
\end{equation}
where $K_0(x)$ is the modified Bessel function. $G_{1}$, $G_{2}$ are the differential conductances
of the quantum dots, without the contributions of the ribbons, taken
at bias voltages $V_{1,2}=E/e$, at $T=0$, and
measured for the normal state of the aluminum film. These conductances should be roughly proportional
to the densities of states in the quantum dots 1 and 2, which, in turn, are sets of Lorentzian peaks centered
around discrete energy levels. Taking $r= 180$ nm, $\xi= 180$ nm, $R_{\Box}=1$ $\Omega$, $G_1=G_2=0.4$ mS, and $T\to 0$,
we estimate the maximum value of the non-local conductance as $g_{12}^{\max}\approx G_{CPS}^{\max}\approx 5$ nS,
which gives the splitting efficiency $\eta\sim 0.1\%$.
%Part of this discrepancy might be related to the titanium layer, which may lead to enhancement of effective $R_{\Box}$.
We believe that the remaining discrepancy may be attributed to the influence of high Ohmic graphene ribbons,
the presence of which can lead to disorder-enhanced crossed Andreev reflection \cite{Golubev2009}.
The point is that Eqs. (\ref{cond}) are derived
neglecting Coulomb blockade in the quantum dots. The ribbons have
large and gate sensitive resistances, and they even host additional small Coulomb blockaded dots near the Dirac point. The theory of Ref. \onlinecite{Golubev2009} predicts
that, in this configuration, $g_{12}$ may increase strongly due to disorder enhanced Andreev scattering.

We have also observed that the value of the splitting efficiency varies by a factor of four between samples with different dot separation
(see Fig. S8 of the supplementary material).
This variation emphasizes the importance of sample geometry. Indeed,
since a diffusive aluminum film is described by Dorokhov distribution of transmission eigenvalues \cite{Dorokhov,Lesovik2011},
there exists fully transmitting, inter-dot channels, which may provide an additional route for enhancing the splitting efficiency.

Since the distance between the junctions is only about 10 times longer than the mean free path, one can justly expect a contribution to CPS from ballistic electron trajectories between the junctions.  Because these trajectories realize one-dimensional ballistic superconductor, the splitting efficiency, according to Ref. \onlinecite{Leijnse2013}, should decay slowly with the distance.  Furthermore, as found recently in Ref. \onlinecite{Sadovskyy2015}, by  tuning the dot resonances properly in the 1d setting, one can achieve even a unitary limit (i.e. 100 \% efficiency).  This unitary limit was found to be extremely robust against variation in the dot parameters (e.g. the barrier transparency), which suggests that our experimental findings may be appreciably influenced by the existence of  ballistic channels. It is noteworthy that in the case of unity CPS probability, $T_{cps}=1$, the interference processes forming the transmission involve precisely two trajectories, like in a Mach-Zehnder interferometer, with both paths of length $\sim \xi_0$. Consequently, the resonance is robust also against a deviation from the pure 1d situation. Indeed, semi-classical trajectories connecting two dots in the experimental configuration do ensure an appreciable probability to join the dots despite the finite spreading of traveling wave packets corresponding to quantum particles.
%Additionally, in a diffusive superconductor, impurities impede the wave packet spreading, thus enhancing the effective one-dimensionality of the setting.

Our measured splitting efficiency vs. dot separation complies favorably with the model of Ref. \onlinecite{Sadovskyy2015}, but the disparity of the conditions for individual points has to be kept in mind (see Fig. S8 in the supplement and the related discussion). According to Ref. \onlinecite{Sadovskyy2015}, in a short superconductor, $L \ll \xi_0$, where the electron-to-hole reflection amplitudes are small $r_{\rm eh(he)}\propto L/\xi_0$, electrons pass through the superconductor freely so that the probability $T_{EC}$ is large, while in a long superconductor, $L\gg\xi_0$, the transmission probability both for electrons and holes through the N-S-N part decays exponentially, $t_{\rm ee(hh)}\propto e^{-L/\xi_0}$. Therefore, a maximum in $T_{cps}$ is obtained around $L \sim \xi_0$ and,up to the natural replacement  of $\xi_0$  by  $\xi $ corresponding to the diffusive case, one can reach a qualitative agreement with  Fig. S8.
 %as alluded by our experimental results.

Previously, high splitting efficiency has been observed in other systems. In the case of CPS in InAs nanowire, the high splitting efficiency was explained by arguing that the Cooper pairs split out from the nanowire with an induced gap underneath the contact \cite{hofstetter2010}; in our case this explanation fails because of the gap between the dots. In NSN structures, crossed Andreev reflection has been observed with normal leads separated by distances comparable to ours \cite{wei2010}.

The advantage of our CPS device configuration compared with the previous quantum-dot-based devices is that we are able to bias the two QDs independently.  By biasing QD1 alone with $V_{ac}= 60$ $\mu$V, we may measure the current in both QDs having the switch A at position 1 and switch B at 4. While sweeping $V_{sg1}$, $g_{1}$ displays conductance oscillation (see Fig. 5(c)). Even though there is no bias voltage on the QD2, the current in QD2 is nonzero at each conductance peak in $g_{1}$. On the left side of the peak in $g_1$, the nonlocal current $I_2$ is positive, while on the right side of the $g_1$ peak, the nonlocal current $I_2$ becomes negative.

The positive nonlocal current $I_2$ is caused by CPS, while the negative one -- by EC. Due to EC, carriers can tunnel from one QD through the superconductor to the other QD, leading to negative nonlocal current on the other side. It is apparent from Eqs. \ref{cond} that if the energy levels of the two QDs are asymmetric, CPS will be favored, and if the energy levels of the two QDs are symmetric, EC will be favored \cite{Feinberg2003, Chevallier2011}. However, so far there is no $direct$ observation of this fact. In a NSN type of structure, CPS and EC could be distinguished by applying a dc bias \cite{russo2005}. Since there were no QDs in the normal metal experiments, it was proposed that the probability of CPS and EC is energy dependent, resulting in a symmetric total effect of CPS and EC with respect to positive and negative bias. In InAs nanowire experiments \cite{hofstetter2011}, the interplay between CPS and EC was observed in relation to the energy level modification of the QD by applied dc bias. The symmetric and asymmetric energy level configurations of the two QDs could not be demonstrated via gate tuning. In our case, we directly observe the CPS and EC while the energy levels of the two QDs are tuned to asymmetric and symmetric configurations without any dc bias.
\begin{figure}[htbp]
\includegraphics[width=0.9\linewidth]{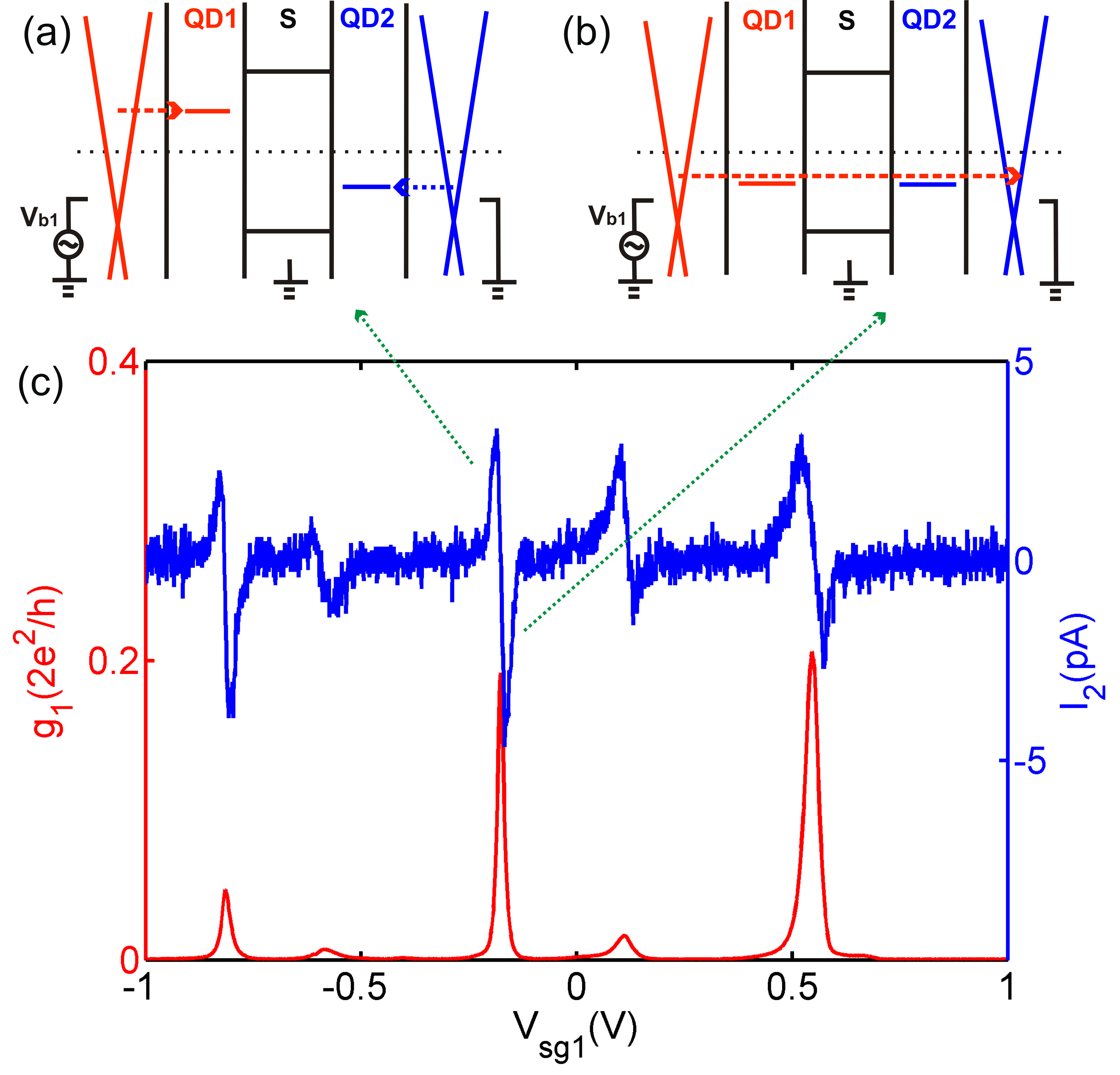}
{\caption{(a) CPS dominates the transport when the engergy levels of the two QDs are asymmetric. (b) EC dominates the transport when the energy levels of the two QDs are symmetric. (c) Conductance $g_1$ and nonlocal current $I_2$ as a function of $V_{sg}$ at $V_{bg}=5$ V with only ac bias on QD1, while the other terminals were shorted to ground.}}
\end{figure}

In Fig. 5(c), being on the left or on right side of the $g_1$ peaks determines the energy level of QD1 to be above or below the Fermi surface. On the left side of the $g_1$ peak, the energy levels of the two QDs are asymmetric (see Fig. 5(a)) and CPS will dominate the transport, leading to positive nonlocal current in $I_2$ (see the blue trace in Fig. 5(c)). On the right side of the $g_1$ peak, the energy levels of the two QDs are symmetric (see Fig. 5(b)) and EC will dominate the transport, leading to negative nonlocal current in $I_2$. The magnitudes of the positive and negative $I_2$ values are of the same order. This indicates that the probability of CPS and EC is roughly equal \cite{Feinberg2003}.

To conclude, we have succeeded in splitting Cooper pairs in graphene by using two spatially separated QDs. We reach about 10 $\%$ CPS efficiency which is much higher than predicted by present theoretical models unless strong enhancement of crossed Andreev reflection by disorder is evoked.  In our work, CPS and EC can be well distinguished from each other by tuning the energy levels of the two QDs to be asymmetric and symmetric with respect to the Fermi level. We find at low bias voltage that the magnitude of CPS and EC  is roughly equal under their respective resonance conditions, which in accordance with theories predicting cancellation of CPS and EC without energy filtering.

We acknowledge fruitful discussions with D. Feinberg, I. Sadovsky, V. Vinokur, and P. Virtanen.  Our work was supported by the Academy of Finland (contracts no. 135908, 278102, and 250280, LTQ CoE) as well as by RFBR Grant No. 14-02-01287. The research leading to these results has received funding from the EU FP7 Programme under grant agreement no 604391 Graphene Flagship, no 323924 iQuOEMS, and the work benefited from the use of the Aalto University Low Temperature Laboratory infrastructure.


\begin{thebibliography}:
\bibitem{lesovik2001} G. B. Lesovik, T. Martin, and G. Blatter, Eur. Phys. J. B \textbf{24}, 287 (2001).
\bibitem{Recher2001} P. Recher, E. V. Sukhorukov, and D. Loss, Phys. Rev. B \textbf{63}, 165314 (2001).
\bibitem{loss1998} D. Loss, and D. P. DiVincenzo,  Phys. Rev. A \textbf{57}, 120 (1998).
\bibitem{beckmann2004} D. Beckmann, H. B. Weber, and H. v. Löhneysen,  Phys. Rev. Lett. \textbf{93}, 197003 (2004).
\bibitem{russo2005} S. Russo, M. Kroug, T. M. Klapwijk, and A. F. Morpurgo, Phys. Rev. Lett. \textbf{95}, 027002 (2005).
\bibitem{cadden2009} P. Cadden-Zimansky, J. Wei and V. Chandrasekhar, Nature Phys. \textbf{5}, 393 (2009).
\bibitem{wei2010} J. Wei and V. Chandrasekhar, Nature Phys. \textbf{6}, 494 (2010).
\bibitem{hofstetter2010} L. Hofstetter, S. Csonka, J. Nyg{\aa}rd and C. Sch\"{o}nenberger, Nature \textbf{461}, 960 (2009).
\bibitem{hofstetter2011} L. Hofstetter, S. Csonka, A. Baumgartner, G. Fulop, S. d’Hollosy, J. Nyg{\aa}rd, and C. Sch\"{o}nenberger, Phys. Rev. Lett. \textbf{107}, 136801 (2011).
\bibitem{Das2012} A. Das, Y. Ronen, M. Heiblum, D. Mahalu, A. V. Kretinin, and H. Shtrikman, Nature Commun.\textbf{3}, 1165 (2012).
\bibitem{herrmann2010} L. G. Herrmann, F. Portier, P. Roche, A. Levy Yeyati, T. Kontos, and C. Strunk, Phys. Rev. Lett. \textbf{104}, 026801 (2010).
\bibitem{schindele2012} J. Schindele, A. Baumgartner, and C. Sch\"{o}nenberger, Phys. Rev. Lett. \textbf{109}, 157002 (2012).
%\bibitem{beenakker} C. W. J. Beenakker, Phys. Rev. Lett. \textbf{97}, 067007 (2006).
\bibitem{cayssol2008} J. Cayssol, Phys. Rev. Lett. \textbf{100}, 147001 (2008).
\bibitem{benjamin2008} C. Benjamin, and J. K. Pachos, Phys. Rev. B \textbf{78}, 235403 (2008).
\bibitem{guttinger2012} J. Guttinger, F. Molitor, C. Stampfer, S. Schnez, A. Jacobsen, S. Droscher, T. Ihn, and K. Ensslin, Rep. Prog. Phys. \textbf{75}, 126502 (2012).
\bibitem{schaffer2012} A. M. Black-Schaffer, Phys. Rev. Lett. \textbf{109}, 197001 (2012).
\bibitem{chamon2012} C. Chamon, C. Y. Hou, C. Mudry, S. Ryu, and L. Santos, Phys. Scr. \textbf{T146}, 014013 (2012).
\bibitem{heersche2007} H. B. Heersche, P. Jarillo-Herrero, J. B. Oostinga, L. M. K. Vandersypen, and A. F. Morpurgo, Nature \textbf{446}, 56 (2007).
\bibitem{dirks2011} T. Dirks, T. L. Hughes, S. Lal, B. Uchoa, Y. F. Chen, C. Chialvo,P. M. Goldbart, and N. Mason, Nature Phys. \textbf{7}, 386 (2011).
\bibitem{Feinberg2003} D. Feinberg, Eur. Phys. J. B \textbf{36}, 419 (2003).
\bibitem{NOTE} The subgap conductance of the large graphene-Al contacts C1 abd C2 are so large that the voltage drop over them can be neglected at small currents.
\bibitem{wiel2003} W. G. van der Wiel, S. De Franceschi, J. M. Elzerman, T. Fujisawa, S. Tarucha, and L. P. Kouwenhoven, Rev. Mod. Phys. \textbf{75}, 1 (2003).
\bibitem{Golubev2009} D.S. Golubev, M.S. Kalenkov, and A.D. Zaikin, Phys. Rev. Lett. \textbf{103}, 067006 (2009).
\bibitem{Dorokhov} O. N. Dorokhov, Solid State Comm. \textbf{51}, 381 (1984).
\bibitem{Lesovik2011} G.B. Lesovik, I.A. Sadovskyy, Phys. Usp. \textbf{54}, 1007 (2011).
\bibitem{Leijnse2013} M. Leijnse and K. Flensberg, Phys. Rev. Lett. \textbf{111}, 060501 (2013).
\bibitem{Sadovskyy2015} I. A. Sadovskyy, G. B. Lesovik, and V. M. Vinokur, to be published.
\bibitem{Chevallier2011} D. Chevallier, J. Rech, T. Jonckheere, and T. Martin, Phys. Rev. B \textbf{83}, 125421 (2011).
\end{thebibliography}
\end{document}